\documentclass[12pt,a4paper]{article}

\usepackage{graphics}
\usepackage{latexsym}
\usepackage{amsmath}
\usepackage{amssymb}

\title{Global Warming: some back-of-the-envelope calculations}
\author{C. Fabara and B. Hoeneisen}
\date{\small{Universidad San Francisco de Quito, \\
14 March 2005}}

\begin{document}
\maketitle

\begin{abstract}
\noindent
We do several simple calculations and measurements in an effort to
gain understanding of global warming and the carbon cycle.
Some conclusions are interesting:
(i) There has been global warming since the end of the
\textquotedblleft{little ice age}" around 1700.
There is no statistically significant evidence of acceleration of 
global warming since 1940.
(ii) The increase of $CO_2$ in the atmosphere,
beginning around 1940, accurately tracks the burning of fossil
fuels.  Burning all of the remaining economically viable reserves of oil, gas and
coal over the next 150 years or so will approximately double the pre-industrial
atmospheric concentration of $CO_2$. The corresponding increase in
the average temperature, due to the greenhouse effect, is quite uncertain:
between 1.3 and 4.8K. This increase of temperature is (partially?) offset by
the increase of aerosols and deforestation.
(iii) Ice core samples indicate that the pre-historic $CO_2$ concentration and
temperature are well correlated.
We conclude that changes in the
temperatures of the oceans are probably the cause
of the changes of pre-historic atmospheric $CO_2$ concentration.
(iv) Data suggests that large volcanic explosions can trigger transitions
from glacial to interglacial climates.
(v) Most of the carbon fixed by photosynthesis in the Amazon basin
returns to the atmosphere due to aerobic decay.
\end{abstract}

\section{Introduction}
We, two non-experts, present several 
\textquotedblleft{back-of-the-envelope}" 
calculations and some simple measurements
related to global warming and the carbon cycle.
Our purpose is to understand which phenomena are
important in determining the temperature of the Earth.

\section{Data}
In this article we will use the following measured 
data.\cite{hard_science, atmosphere}
The power of the radiation of the Sun
per square perpendicular meter above the atmosphere
is measured to be $I_{Sun} = 1368$W/m$^2$.
\footnote{The power per unit area radiated by a black
body at absolute temperature $T$ is $\sigma T^4$,
with $\sigma = 5.67 \cdot 10^{-8}$Wm$^{-2}$K$^{-4}$.
The power radiated by the Sun is
$P = 4 \pi R_{Sun}^2 \sigma T_{Sun}^4$.
The power per unit perpendicular area received by Earth
(outside of the atmosphere) is 
$1368$W/m$^2 = P/( 4 \pi R_{\textrm{Sun-Earth}}^2)$. From these
equations we obtain the effective temperature of 
sun light: $T_{Sun} = 5780$K.}
This is the
\textquotedblleft{solar constant}". Since the
surface of the Earth is 4 times the surface of
a disk of the same radius, the average incident solar
power per square meter of the Earth's surface is
$1368/4 = 342$W/m$^2$.
All fractions listed below refer to $342$W/m$^2$.
The fraction of sun light that is reflected by the Earth
is measured to be 0.31 (0.21 by clouds, 0.06 by air
including dust and water vapor,
and 0.04 by the ground). 
This is the \textquotedblleft{albedo}"
of the Earth. 
The remaining fraction (0.69) is absorbed
by the Earth ($168$W/m$^2$ by the surface, $48$W/m$^2$ by the troposphere
(water vapor and aerosols), $10$W/m$^2$ by the stratosphere, 
and $10$W/m$^2$ by clouds).
So, the net solar power absorbed by the Earth is
$342 \cdot 0.69 = 236$W/m$^2$ for average cloud
coverage.

The infrared radiation of the Earth to space measured by
satellites is $235$W/m$^2$ for average cloud
coverage ($88$W/m$^2$ by clouds, $126$W/m$^2$ by water vapor and
$CO_2$, and $21$W/m$^2$ by the surface through the 
infrared atmospheric window).
So, the incoming and outgoing
energies balance to high accuracy and determine
the temperature of the biosphere.

The average temperature of the surface of the Earth
is 288K$ = 15^o$C. The average temperature of the
atmosphere is about 250K$ = -23^o$C.

In cloud free conditions the average solar power absorbed
is $286$W/m$^2$, while the average emitted terrestrial 
radiation is $266$W/m$^2$. 
The corresponding numbers for average cloud cover are
$236$W/m$^2$ and $235$W/m$^2$ as mentioned before.
Note that decreasing the cloud
cover has a net heating effect.

\begin{figure}
\begin{center}
\scalebox{0.9}
{\includegraphics{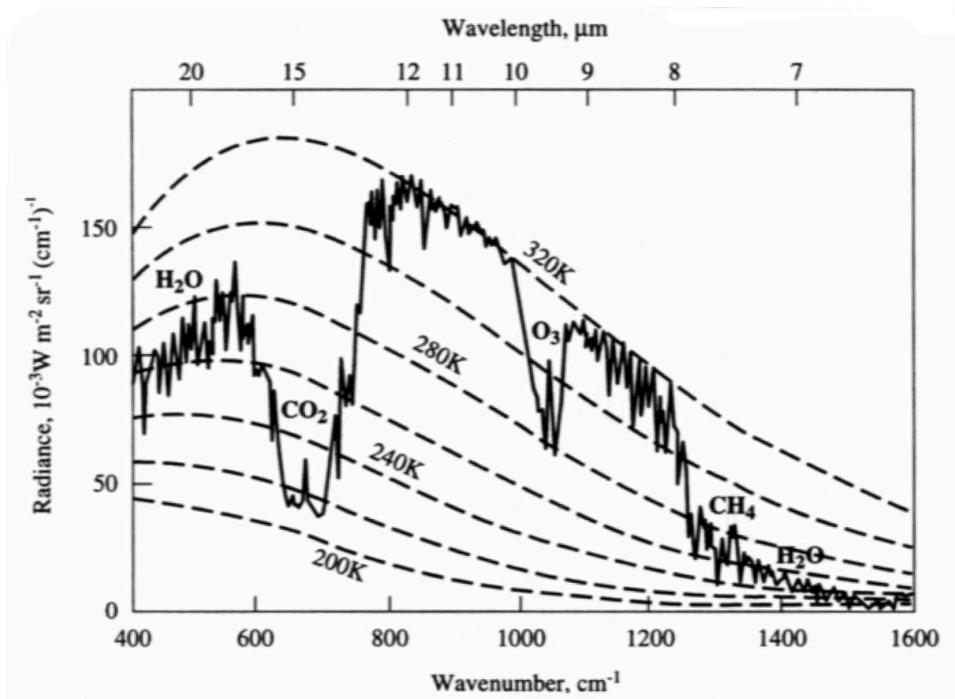}}
\vspace*{-0.3cm}
\caption{Radiation spectrum of the Earth as observed by a satellite
over Africa.\cite{earthrad} Wavenumber is $1/\lambda$. Blackbody spectra
at several temperatures are also shown. Multiply the numbers on the vertical axis
by $\pi$ to obtain the radiation in $10^{-3}$W m$^{-2}($cm$^{-1})^{-1}$ (with
m$^{2}$ now referring to the surface of the Earth instead of the 
satellite antenna).}
\label{earthradiation}
\end{center}
\end{figure}

The Earth radiation spectrum measured from a satellite (Figure \ref{earthradiation})
shows that the infrared windows ($\lambda = 8$ to $9\mu$m
and 10 to $13\mu$m) radiate as a blackbody at the surface temperature,
the $CO_2$ absorption band ($\lambda = 14$ to $16\mu$m) radiates
at 215K (corresponding to the altitude of the tropopause at 12km),
and the $H_2O$ bands ($\lambda < 8\mu$m and
$\lambda > 17\mu$m) radiate at $\approx 260$K (corresponding to an altitude of 5km).

\section{A simple model}
Let us begin with a simple model. From observations we know that, 
to high accuracy, there is
an equilibrium between incident solar power and radiated
infrared power. Therefore
\begin{equation}
\frac{1}{4} \epsilon_v I_{Sun} = \epsilon_{IR} \sigma T^4,
\label{T}
\end{equation}
where $I_{Sun} = 1368$W/m$^2$ is the solar constant,
$\epsilon_v$ is the fraction of incident sun light power
that is absorbed by the Earth, \textit{i.e.} it is the
\textquotedblleft{emissivity in the visible}",
$\epsilon_{IR}$ is the emissivity in the infrared,
$\sigma = 5.67 \times 10^{-8}$Wm$^{-2}$K$^{-4}$ is the
Stefan-Boltzmann constant, and $T$ is the temperature of
the surface of the Earth in degrees Kelvin.
The factor $\frac{1}{4}$ was explained in Section 2.

We note that if $\epsilon_v = \epsilon_{IR}$ the
equilibrium temperature is 279K, quite close to the
observed mean temperature of 288K.

From the data presented in Section 2 we conclude that
$\epsilon_v \approx 0.69$ for the Earth as a whole,
$\epsilon_v \approx 0.55$ for areas covered by clouds
(it varies from 0.6 for cirrus to 0.1 for 
cumulonimbus), 
$\epsilon_v \approx 0.9$ for cloudless ground,
and $\epsilon_v \approx 0.2$ for snow. Because
snow is mostly at high latitude we sometimes replace
$\epsilon_v \approx 0.1$ for snow.

From the data presented in Section 2 we estimate that
$\epsilon_{IR} \approx 0.605$ for the Earth as a whole,
$\epsilon_{IR} \approx 0.56$ for areas covered by clouds,
$\epsilon_{IR} \approx 0.69$ for cloudless ground,
and $\epsilon_{IR} \approx 0.3$ for snow. 

These \textquotedblleft{effective}" emissivities are valid
for the model of this Section, \textit{i.e.} an Earth surface
characterized by a single temperature, and
includes the atmosphere with its greenhouse gases.
The model is too crude to account for all observations, so these
effective emissivities are approximate.
A summary of effective emissivities is presented in Table \ref{emissivities}.
In the last column of the table we show the equilibrium
temperature for 100\% coverage of black body, cloudless ground,
clouds or snow. The last row shows the world average.

\begin{table}
\begin{center}
\begin{tabular}{|c|c|c|c|}
\hline
 & infrared & visible & $T$ for 100\% cover \\
\hline
black body & 1.00 & 1.00 & 279K \\
cloudless ground & $\approx 0.69$ & 0.90 & 298K \\
clouds & $\approx 0.56$ & 0.55 & 278K \\
snow & $\approx 0.30$ & 0.10 (0.20) & 212K (252K) \\
\hline
whole Earth & 0.605 & 0.69 & 288K \\
\hline
\end{tabular}
\end{center}
\caption{Effective emissivities for infrared and visible radiation.
The last column shows the temperature corresponding to a 100\%
cover of black body, cloudless ground, cloud or snow. 
The last row corresponds to 50\% cloud cover, 5\% snow
cover, leaving 45\% clear ground.
The measured average temperature is 288K.}
\label{emissivities}
\end{table}

\section{Two box model}
\begin{figure}
\begin{center}
\scalebox{0.6}
{\includegraphics{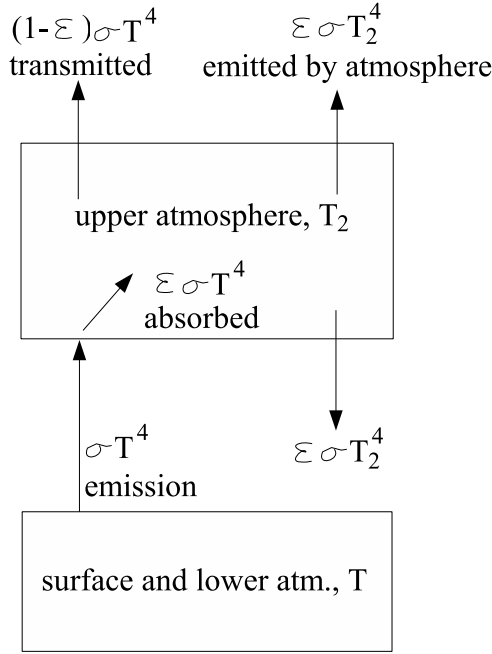}}
\vspace*{-0.3cm}
\caption{Two-box model of the greenhouse effect.\cite{hard_science}}
\label{greenhouse}
\end{center}
\end{figure}

Figure \ref{greenhouse} shows two boxes: one for the
upper atmosphere at temperature $T_2$, and one for the
surface and lower atmosphere at temperature $T$.\cite{hard_science} The surface
radiates $\approx \sigma T^4$. A fraction $1 - \epsilon$
of this radiation gets through the upper atmosphere
(this is the infrared window), and a fraction $\epsilon$
is absorbed by greenhouse gases and aerosols in
the atmosphere. The upper atmosphere
radiates $\epsilon \sigma T_2^4$ back to the lower atmosphere, and
$\epsilon \sigma T_2^4$ out to space. The total radiation
out to space is
\begin{equation}
\sigma T^4 - \epsilon \sigma \left( T^4 - T_2^4 \right).
\label{to_space}
\end{equation}
The last term reduces the infrared emission of the Earth (at a given $T$)
because the upper atmosphere is colder than the surface.
This last term is the greenhouse effect. 
The effective emissivity is
\begin{equation}
\epsilon_{IR} = 1 - \epsilon \left[ 1 - \left( \frac{T_2}{T} \right)^4 \right].
\label{surface_emissivity}
\end{equation}
We estimate $\epsilon \approx 0.813$ (see next section). 
For $T = 288$K we obtain $T_2 \approx 244$K.
Note that heating of the upper atmosphere reduces the greenhouse effect.
Adding greenhouse gases to the atmosphere has two opposing
effects on $T$: $\epsilon$ increases and $T_2$ increases.

\begin{figure}
\begin{center}
\scalebox{0.6}
{\includegraphics{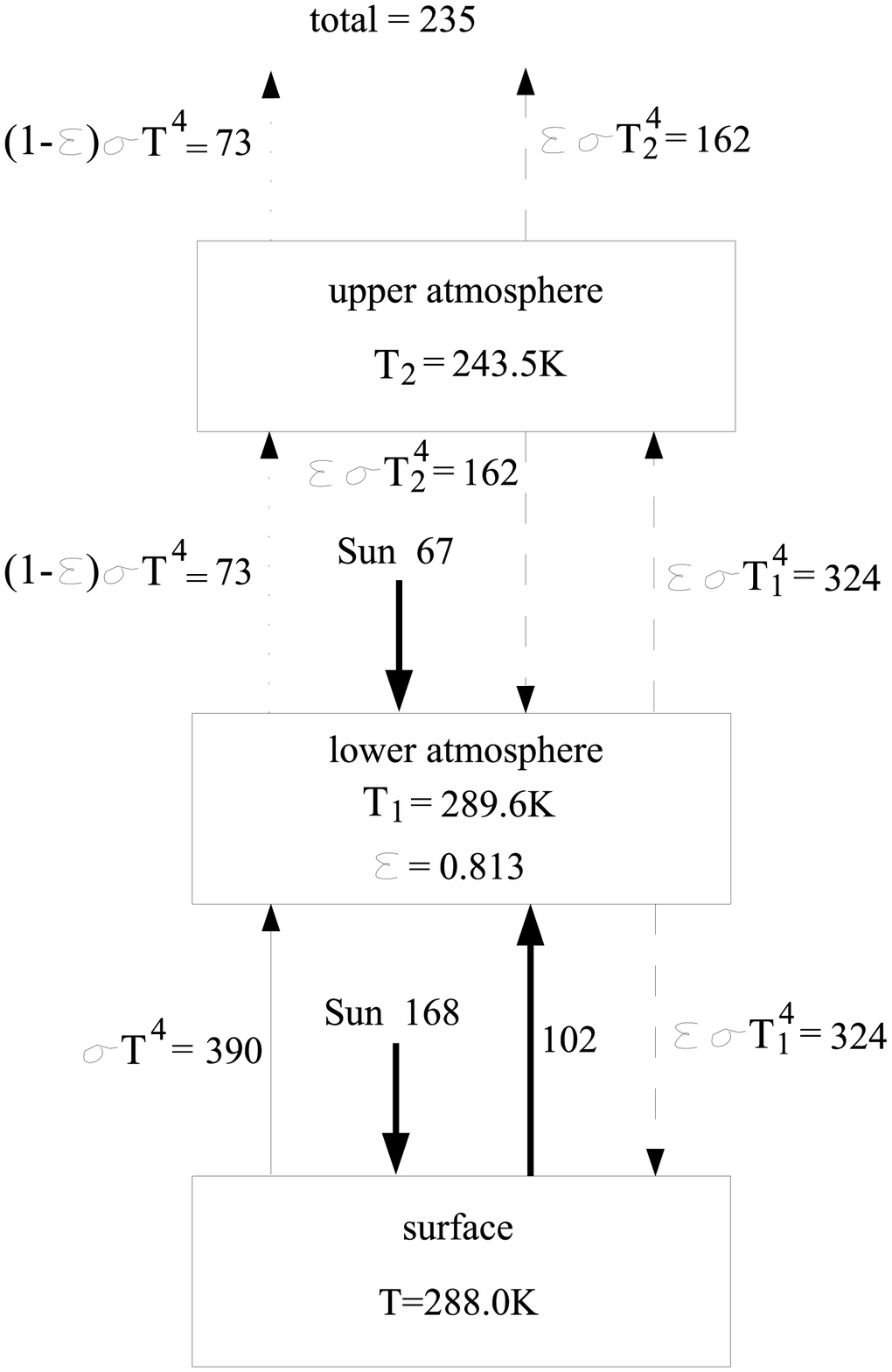}}
\vspace*{-0.3cm}
\caption{Three-box model of the greenhouse effect. The numbers
are in units of W/m$^2$. Dotted lines correspond to the
infrared window, while dashed lines correspond to frequencies absorbed
or emitted by greenhouse gases in the atmosphere (mainly $H_2O$, $CO_2$ and ozone).
Thermals, evaporation and transpiration are estimated to transport $102$W/m$^2$.}
\label{threebox}
\end{center}
\end{figure}

\section{Three box model}
The three box model is
shown in Figure \ref{threebox}. 
The boxes are the surface of the Earth at temperature $T$,
the lower atmosphere at temperature $T_1$, and the upper atmosphere
at temperature $T_2$. 
The total radiation to space is equal to the total solar radiation
absorbed ($235$W/m$^2$), which we assume independent of the concentration
of $CO_2$. 
The solar power absorbed per unit area is
$168$W/m$^2$ by the surface, and $67$W/m$^2$ by the atmosphere.
Thermals, evaporation and transpiration are estimated to transport $102$W/m$^2$.\cite{hard_science}
The atmosphere absorbs radiation at frequencies near vibration resonances of
the greenhouse gases, mainly $H_2O$, $CO_2$ and ozone. The fraction of infrared radiation
absorbed is $\epsilon$. The atmosphere also radiates at these 
frequencies. The emissivity of the surface in the infrared is taken to be 1.
The sum of powers entering each box is zero.
The greenhouse effect is due to the low temperature of the upper atmosphere,
so that emission to space is reduced at frequencies near resonances as shown 
in Figure \ref{earthradiation}.

\begin{table}
\begin{center}
\begin{tabular}{|c|c|c|c|c|}
\hline
$\epsilon$ & $T$ & $T_1$ & $T_2$ & comment \\
\hline
0.0 & 253.72K & n.a. & n.a. & 235W/m$^2$ radiation, no atmosphere \\
0.05 & 238.94K & 452.5K & 380.5K & \\
0.3 & 250.51K & 310.1K & 260.8K & \\
0.5 & 262.20K & 290.5K & 244.3K & model breaks down for $\epsilon < 0.5$ \\
0.6 & 269.20K & 287.3K & 241.6K & \\
0.7 & 277.24K & 286.9K & 241.3K & \\
0.8 & 286.65K & 289.1K & 243.1K & \\
0.813 & 288.00K & 289.6K & 243.5K & 290ppm $CO_2$ \\
0.833 & 290.14K & 290.4K & 244.2K & 580ppm $CO_2$ \\
0.9 & 297.91K & 293.9K & 247.1K & \\
1.0 & 311.81K & 301.7K & 253.7K & opaque atmosphere \\
\hline
\end{tabular}
\end{center}
\caption{Temperatures of the surface ($T$), lower atmosphere ($T_1$), and
upper atmosphere ($T_2$) in the three box model as a function of the
emissivity of the atmosphere in the infrared ($\epsilon$).
The absorbed Sun power and power of thermals, evaporation and transpiration
are assumed to be constant (independent of the concentration of $CO_2$).}
\label{epsilon-T}
\end{table}

In Table \ref{epsilon-T} we show the temperatures as a function of
the emissivity of the atmosphere in the infrared ($\epsilon$).
The absorbed sun power and power of thermals, evaporation and transpiration
are assumed to be constant (independent of the concentration of $CO_2$),
so the model breaks down for $\epsilon \le 0.5$ (with no atmosphere
we should obtain $T = 253.72$K for 235W/m$^2$ black-body radiation).

It has been estimated that doubling the concentration of $CO_2$ 
from (290ppm to 580ppm) decreases the radiation to space 
by $3.75 \pm 0.25$W/m$^2$\cite{hard_science}
(before temperatures are allowed to adjust)
which corresponds to increasing $\epsilon$ from 0.813 to 0.833.
Note, in Table \ref{epsilon-T}, that doubling the concentration of $CO_2$ raises the
surface temperature $T$ by 2.1K.
Do we trust this result? No! Most of the radiation to space is from the
upper atmosphere. The temperature of the upper atmosphere may change with
$CO_2$ concentration due to energy fluxes or absorptions
not considered in this simple model. 
Also the solar inputs and power of thermals, evaporation and transpiration
surely depend directly or indirectly on the $CO_2$ concentration
(by changing $T$ and $T_1$).

\section{Global warming from Earth emission spectra}
Let us try a different approach that relies only on the Earth emission spectra
shown in Figure \ref{earthradiation}. We estimate the increase in
surface temperature due to a doubling of the concentration of $CO_2$
(from 290ppm to 580ppm). We take the average Earth temperature to be 288K
(instead of the 320K of Africa shown in Figure \ref{earthradiation}).

$CO_2$ has two absorption bands in the tail of the solar spectrum, which 
absorb $\approx 0.8$W/m$^2$. We will neglect this absorption (a correction
could be applied later). Increasing the concentration of $CO_2$
will widen the absorption resonance seen in Figure \ref{earthradiation}.
How will the spectra respond? If the Earth albedo remains constant, then the
area below the spectra in  Figure \ref{earthradiation} will
remain constant. Then the spectra will respond by
increasing (or in general, modifying) the surface temperature, and/or 
the water emission temperature ($T_{H_2O} \approx 260$K), 
and/or the $CO_2$ emission temperature ($T_{CO_2} \approx 215$K).
Table \ref{epsilon-T} suggests that $T_1$ and $T_2$ vary less than the 
surface temperature $T$. So, in this section, without much justification,
we will assume that (i) $T_{H_2O}$ and $T_{CO_2}$ remain constant, and
(ii) the Earth albedo remains constant. So, in this approximation, the
only response to the widening of the $CO_2$ absorption resonance is to
increase the surface temperature to conserve the area under the 
spectrum in Figure \ref{earthradiation}.

Let us estimate the widening of the $CO_2$ resonance using only the data in
Figure \ref{earthradiation}. This widening in W/m$^2$ (before any temperature
has a time to react) is
called \textquotedblleft{radiative forcing}". 
We label points of transmittance equal to 1, 0.7, 0.5, 0.25, 0.06, 
0.004 and 0.0
along the side slopes of the absorption resonance,
and then shift them downward (at constant wave number).
\footnote{To find where to place the points and by how much 
they shift downward, we simulated a ten-layer atmosphere 
model on a spreadsheet.}
The result is a radiative forcing 
$\approx 4$W/m$^2$.
The corresponding change in surface temperature, in responce to a 
doubling of the concentration of $CO_2$, with the assumptions discussed above,
is $\approx 2.5$K.

Another estimate would let $T$ and $T_{H_2O}$ vary by the same amount, 
while $T_{CO_2}$ remains constant. The corresponding warming is
$\approx 1.3$K.

\section{Detailed models}
The results presented in this section were obtained from \cite{hard_science}.
The surface and troposphere are tightly coupled 
(by non-radiative heat exchanges), while the
coupling between the troposphere and stratosphere is
relatively weak.
For this reason changes in
the surface temperature are largely determined by changes
in the net (incoming minus outgoing) radiation at the
tropopause (at an altitude between 10 and 20 km).
This change in net radiation at the tropopause per unit surface
area (before any temperatures are allowed to change), $\Delta R$, is called
\textquotedblleft{radiative forcing}". The change in surface
temperature is expressed as $\Delta T = \Delta R / \lambda$,
where $\lambda$ is the \textquotedblleft{radiative damping}".
If the Earth were a blackbody (radiating the same amount as the Earth),
then $\lambda = 3.7$Wm$^{-2}$K$^{-1}$. However due to various
feedback effects, it is estimated, using detailed models, that
$\lambda = 2.0 \pm 0.5$Wm$^{-2}$K$^{-1}$.\cite{hard_science}
So the net feedbacks are positive.
\footnote{Why $\lambda$ drops can be understood from our
discussion in Section 6.}

The change in radiative forcing and surface temperature, 
due to changes in composition
of the atmosphere since the pre-industrial period 
until 1995, is estimated by detailed models\cite{hard_science}
to be $\Delta R = 2.6 \pm 0.6$Wm$^{-2}$ 
and $\Delta T = 1.3 \pm 0.44$K due to all greenhouse
gases (about half of this is due to $P_{CO_2}$ increasing
from 290ppm to 350ppm), and
$\Delta R = -1.9 \pm 1.1$Wm$^{-2}$ 
and $\Delta T = -1.0 \pm 0.6$K due to all aerosols
(we have added errors in quadrature). 
Deforestation changes the albedo of the Earth, causing a 
change of temperature $\approx -0.4$K.\cite{deforestation}
So we do not know wether the increase of greenhouse gases,
aerosols and deforestation has a net heating or cooling effect.

Let us now consider a doubling of the pre-industrial concentration
of $CO_2$ (from 290ppm to 580ppm). The change in 
radiative forcing is $\Delta R = 3.75 \pm 0.25$Wm$^{-2}$\cite{hard_science}
and the corresponding change of surface temperature is
$\Delta T = 1.9 \pm 0.5$K.
This result, obtained from detailed models\cite{hard_science}
is in agreement with paleo-climate and historical data
(and with our back-of-the-envelope estimates of Sections 5 and 6).
In comparison, the change in temperature since the 
\textquotedblleft{little-ice-age}" (1700) to the present (1995) has been
$\approx 1.2$K.\cite{global-warming}
It is worth mentioning that various detailed models obtain
$\Delta T$ in the range from 1.3K to 4.8K for doubling of
$CO_2$\cite{hard_science}. 
The range of results is large so
this is a difficult problem that does not seem to converge:
as the model becomes more complex, the uncertainties appear to
increase!

\section{Ice ages}
For given emissivities, the Earth temperature is stable
because an increase in temperature causes an increase
in the infrared radiation, which in turn brings the temperature
back to its equilibrium value. 
However, the equilibrium 
temperature depends on the emissivities. For example,
a world-wide snow storm could result in a shift
of the equilibrium temperature from 288K to as low as 252K,
see Table \ref{emissivities}, which would prevent
snow from melting. In this way an ice age might begin.

The mean annual temperature averaged over the Earth fluctuates
from year to year with a standard deviation of about
$\approx 0.22^0$C (during the 20'th century). 
Inter-glacial periods last typically 
$\approx 50$ thousand years. With 50\% probability we expect
a fluctuation of at least 4.2 standard deviations in 
50000 tries. This corresponds to $\approx 1^0$C.
Is such a world-wide temperature fluctuation sufficient
to throw us into an ice age?

\begin{figure}
\begin{center}
\scalebox{0.6}
{\includegraphics{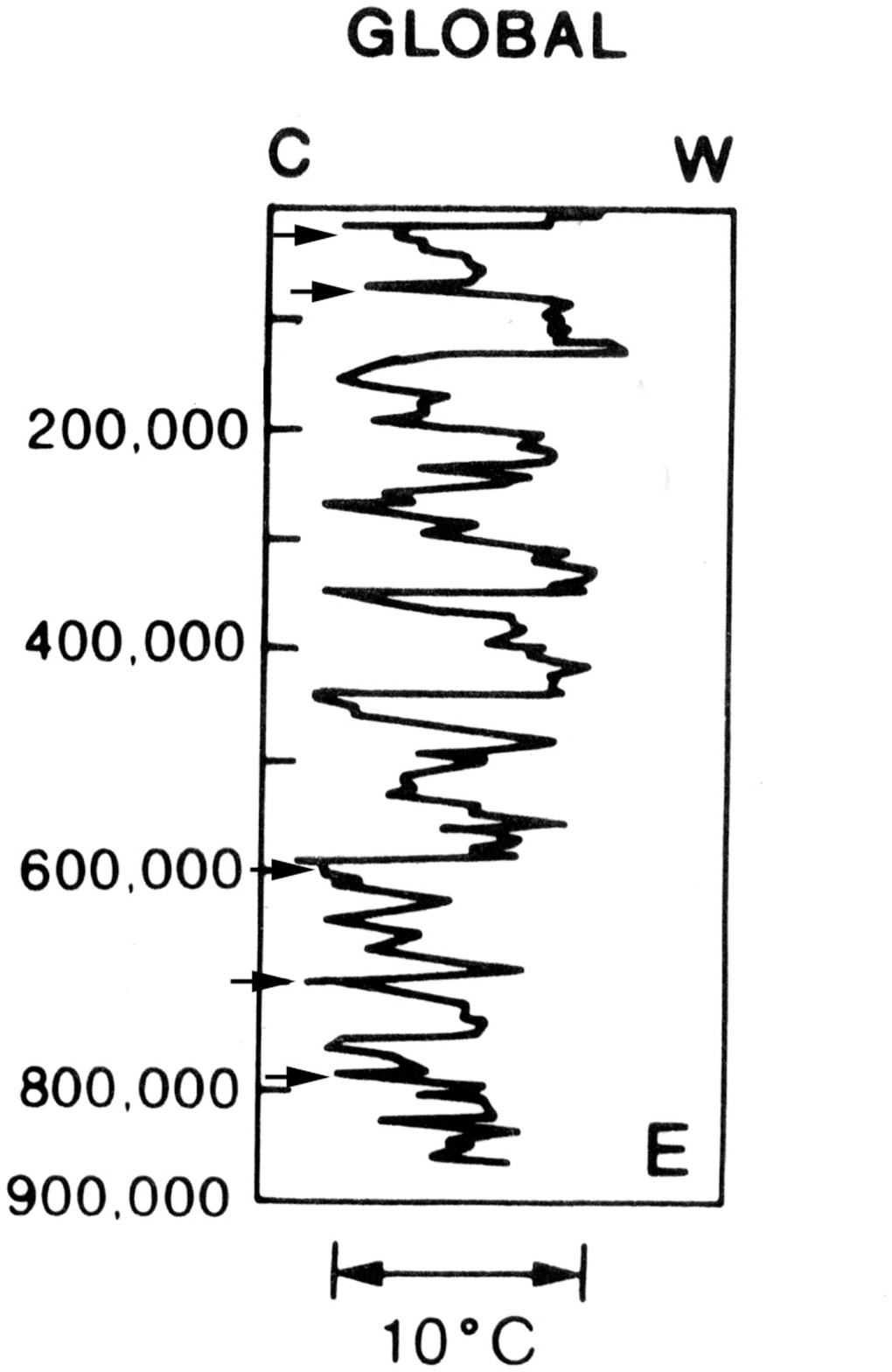}}
\vspace*{-0.3cm}
\caption{Ice-core temperature data of the last 900 thousand years\cite{atmosphere}, 
compared with the largest known volcanic explosions in that period 
(arrows)\cite{volcano}.}
\label{icecore}
\end{center}
\end{figure}

To get out of an ice age, an event is needed to
change the emissivities, perhaps volcanic eruptions
(or meteorite impacts) that cover snow with ash.
Table \ref{volcano_eruptions} is a list of the largest
known volcanic eruptions in the last million years.
Each one of them (except perhaps Taupo in isolated and tropical New Zealand) 
corresponds (within errors) to a transition from a 
glacial to an interglacial period, see Figure \ref{icecore}. 
It is interesting to note that an explosion that ejects
$2\times10^{15}$kg corresponds to an average of
4kg$\cdot$m$^{-2}$ over the entire Earth, so that such an
event could have a sizable effect on the albedo.
If the volcanic
origin of the transitions is correct, we should find
evidence for other large eruptions at approximately
770, 430, 340, 270 and 140 thousand years ago. 
Note however that volcanic explosions have a short-term 
cooling effect.\cite{deforestation}

On 4 November 2002 the volcano El Reventador 
in Ecuador exploded and covered
Cayambe with ash. Two years later we see a barren Cayambe with
large patches of glacier gone. 
Similarly, volcanic ash from Tungurahua has been
falling on Chimborazo for the last few years.
Chimborazo also looks barren compared with what it used to be.

\begin{table}
\begin{center}
\begin{tabular}{|c|c|c|}
\hline
Caldera name & ejected mass [$10^{15}$kg] & date [$10^3$ years ago] \\
\hline
Toba (Indonesia) & 6.9 & 74 \\
Yellowstone (Wyoming, USA) & 2.2 & 600 \\
Porsea (Toba, Indonesia) & 2.0 & 790 \\
Taupo (New Zealand) & 1.3 & 26.5 \\
Long Valley (California, USA) & 1.2 & 700 \\
\hline
\end{tabular}
\end{center}
\caption{Large explosive volcanic eruptions in the last
million years.\cite{volcano}}
\label{volcano_eruptions}
\end{table}

\section{Width of the $15\mu$m line of $CO_2$.}
The $CO_2$ molecule is linear, and has three normal
modes of oscillation. They have wavelengths of
$15\mu$m, 7.46$\mu$m and 4.26$\mu$m. The mode with
wavelength 7.46$\mu$m has no dipole moment and does not
couple to electromagnetic radiation. The mode
with wavelength 4.26$\mu$m barely overlaps the
black-body radiation spectrum of the Earth.
So the important spectral line has a wavelength 
$\lambda = 15\mu$m. We calculate the half-width of this
line (neglecting molecule collisions) at the points of half power to be
$\Delta \lambda / \lambda \approx 10^{-13}$.

Let us consider molecule collisions. The calculated 
mean free path at a pressure of one atmosphere
is $l \approx 1\mu$m, the rms thermal velocity
of $CO_2$ is $v = 400$m/s at 300K, and the mean time between collisions
is $\tau \approx 2.5$ns. If the time between collisions is $t$, then
the half-width of the spectral line between points of half power
is $\Delta \lambda / \lambda \approx 0.45 \lambda/(c t)$.
For $t = 2.5$ns the result is
$\Delta \lambda / \lambda \approx 10^{-5}$.
The probability that $t$ is in the interval $dt$ is
$\tau^{-1} \exp[-t/\tau]\cdot dt$. 
So the spectral line has very long tails that 
absorb electromagnetic radiation.
These long tails explain the wide absorption band of $CO_2$
seen by satellites.

We have calculated the half-width $\Delta \lambda$ of the window
corresponding to the $\lambda = 15\mu$m line of $CO_2$
at which the power per unit area getting through the atmosphere
drops by a factor $e^{-2}$:
\begin{equation}
\Delta \lambda \approx \frac{\lambda^2 q}{4 \pi c}
\left[ 
\frac{2 P_{CO_2} L \eta_0}{k T m \tau}
\right]^{\frac{1}{2}},
\label{window}
\end{equation}
where $\eta_0 = 120 \pi$ Ohm is the impedance of free space,
$P_{CO_2}$ is the partial pressure of $CO_2$ at sea
level, $L = kT/(Mg) = 5550$m is the altitude at which the partial pressure of
$CO_2$ drops by a factor $e^{-1}$,
$c$ is the velocity of light, $m$ is the reduced mass of 
a carbon atom oscillating against two oxygen atoms, $T$ is the
absolute temperature, $\tau$ is the mean time between collisions
of $CO_2$ molecules, and $q$ is the effective charge of the carbon in $CO_2$.
Substituting numerical values we obtain 
$\Delta \lambda \approx 0.6\mu$m for 300ppm $CO_2$,
in reasonable agreement with
$\approx 1.0\mu$m measured by satellites.
So we understand the long tails of the spectral line of
$CO_2$. Note that doubling $P_{CO_2}$ increases $\Delta \lambda$
by a factor $\sqrt{2}$.
\footnote{Increasing $\Delta \lambda$
by a factor $\sqrt{2}$ results in a radiative forcing greater than
our estimate of Section 6, so the calculation should be taken with a 
grain of salt when applied so far out on the tails of the resonance.}

\begin{table}
\begin{center}
\begin{tabular}{|c|c|c|c|c|c|c|}
\hline
pH & $CO_2$ & $CO_2$ & $HCO_3^-$ & $HCO_3^-$ & Gt of C & Gt of C \\
\hline
   & $0^0$C & $25^0$C & $0^0$C & $25^0$C & $0^0$C & $25^0$C \\
\hline
7.0 & $2.49 \cdot 10^{-4}$ & $1.04 \cdot 10^{-4}$ & $2.12 \cdot 10^{-3}$ & $1.43 \cdot 10^{-3}$ & 843 & 550 \\
7.4 & $2.49 \cdot 10^{-4}$ & $1.04 \cdot 10^{-4}$ & $5.32 \cdot 10^{-3}$ & $3.60 \cdot 10^{-3}$ & 1994 & 1325 \\
8.3 & $2.49 \cdot 10^{-4}$ & $1.04 \cdot 10^{-4}$ & $4.23 \cdot 10^{-2}$ & $2.86 \cdot 10^{-2}$ & 15224 & 10270 \\
\hline
\end{tabular}
\end{center}
\caption{The columns contain the ocean pH; the number of kilograms of carbon
per cubic meter of ocean at $0^0$C and $25^0$C in the form
$CO_2$ and $HCO_3^-$\cite{CO2}; and the corresponding total carbon in the
upper 1000m of the oceans. 
The calculations correspond to 300ppm of $CO_2$ in the atmosphere.
In comparison, the atmosphere with
300ppm of $CO_2$ contains 430Gt of carbon. Oceans have a pH
between 7.4 and 8.3.\cite{CO2}}
\label{oceanC}
\end{table}

\section{Equilibrium of $CO_2$ between atmosphere and oceans}
The chemistry of carbon in the oceans is beautifully described in
\cite{CO2}. We are interested in these two equations of chemical
equilibrium:
\begin{equation}
[CO_2] = K_H P_{CO_2},
\label{Henry}
\end{equation}
which describes the solution of $CO_2$ in sea water, and
\begin{equation}
[H^+] [HCO_3^-] = K_{a1} [CO_2]
\label{bicarbonate}
\end{equation}
that describes the ionization of hydrated carbon dioxide.
$P_{CO_2}$ is the partial pressure of $CO_2$ in the atmosphere
(in atmospheres), and $[...]$ are concentrations in sea water
(in mol per liter). The equilibrium constants have these values
at $0^0$C ($25^0$C): $-\log K_H = 1.16$ ($1.54$), and
$-\log K_{a1} = 6.07$ ($5.86$) for 0.7 ionic strength.\cite{CO2}
The calculations shown in Table \ref{oceanC} 
assume $P_{CO_2} = 0.0003$atm (\textit{i.e.}300ppm).

Assuming that the total carbon in the air and the top 1000m of the oceans
is constant, and that 300ppm corresponds to $10^0$C and pH = 7.4 (8.3),
we obtain that a $10^0$C increase in the sea water temperature
produces an increase of $CO_2$ in the atmosphere of
40ppm (48ppm).
In comparison,
from Antarctic and Greenland ice cores, it is found that $9.5^0$C changes in
temperature have been accompanied by $\approx 75$ppm change in $P_{CO_2}$.

In conclusion, we qualitatively understand the
coupling of the pre-historic temperature with the atmospheric concentration of
$CO_2$: an increase of ocean temperature reduces its solubility to
$CO_2$, \textbf{causing} an increase of the atmospheric concentration of $CO_2$.
The converse is not true: an increase of $75$ppm $CO_2$ would increase
the temperature, due to the greenhouse effect, by only $\approx 0.8^0$C.

Let us now estimate the effect of acid rain due to a large
volcanic explosion.
Assume that $2\times10^{15}$kg are ejected.
Assume that 0.1\% of this is sulfur (in any chemical form) 
that ends up as sulphuric acid in the oceans. The 
concentration of $H_2SO_4$ in the top 1000m of the oceans
would be $1.8 \cdot 10^{-4}$mmol/liter. We measured the pH
of ocean water off the coast of Ecuador as a function of
$H_2SO_4$ concentration: 7.3, 7.0, 6.3, and 5.5
for 0, 0.0084, 0.0211 and 0.0422mmol/liter respectively.
\footnote{Measurement done several days later on a sample of water taken from
the beach at Atacames. 
A measurement done locally in Mompiche
yielded pH = 7.7.}
So the change of pH is negligible ($\approx 0.006$), and
the corresponding release of $CO_2$ by the oceans is also
negligible.

\section{A comment on the carbon cycle}
It is observed that the concentration of $CO_2$ has increased 
linearly from 325ppm in 1970 to 375ppm in 2004.\cite{maunaloa}
If this increase continues at constant rate it will take 
about $\approx 140$ years to reach twice the pre-industrial concentration of
290ppm.

It is observed that the buildup of atmospheric $P_{CO_2}$ accurately
follows the human emissions of $CO_2$ (due to burning of fossil fuels
and the clearing of land for agricultural and urban use).\cite{hard_science}
The factor of proportionality is observed to be 65\%.
The remaining 35\% of emitted carbon apparently ends up in the
surface (mixed) layer of the oceans and in bio-mass.
No saturation is observed so far, so the time constant with
which carbon reaches the deep oceans is in excess of 50 years
(it is estimated that there are several time constants, some as long
as 2000 years).\cite{hard_science}

There are enough world reserves of fossil fuels (gas, oil and coal) to last
$\approx 150$ years at the present rate of consumption
(6.5Gt C per year).
Burning these reserves would increase $CO_2$ to three times
the pre-industrial concentration assuming 65\% of the emitted
$CO_2$ ends up in the atmosphere. However, on these longer time scales
it is estimated that 30\% (instead of 65\%) of the emitted
$CO_2$ will remain in the atmosphere.\cite{hard_science}
Thus we estimate that burning all of the remaining reserves of
fossil fuels in the next 150 years or so, will result in twice
the pre-industrial $CO_2$ concentration.

\section{Time constant to reach $CO_2$ equilibrium with
the oceans.}
The solubility of $CO_2$ in the oceans decreases with
increasing temperature. Consider a drop in the temperature of
the oceans.
With what time constant does the atmospheric
concentration of $CO_2$ drop due to the increased absorption by
the oceans? Here we assume that every $CO_2$ molecule
colliding with the ocean surface \textquotedblleft{sticks}"
to the surface. The number of $CO_2$ molecules colliding
with a square meter of ocean surface per second is\cite{thermalphys}
\begin{eqnarray}
R = \sqrt{\frac{kT}{2\pi M}} \frac{P_{CO_2}}{kT} = 7\cdot10^{23}
\textrm{m}^{-2}\textrm{s}^{-1}.
\end{eqnarray}
The number of $CO_2$ molecules in the atmosphere per square meter
is $P_{CO_2}/(Mg) = 4\cdot10^{25}$m$^{-2}$. So the time
constant for absorption of $CO_2$ by very cold oceans is
$\approx 90$s (we have taken account that 0.7 of the
surface of the Earth is covered by oceans).
So equilibrium of $CO_2$ and the oceans is very short
if we neglect diffusion of $CO_2$ through the atmosphere.

Let us now consider diffusion.
The partial pressure of $CO_2$ decreases with altitude
exponentially with a characteristic height 
$L = kT/(Mg) = 5550$m. How long does it take a molecule of
$CO_2$ to diffuse 5550m through the atmosphere?
The result is $t = L^2/(vl) \approx 2400$ years.
This result assumes no convection.
Convection will reduce this time considerably.
In fact, the flux of carbon from the atmosphere
to the oceans and \textit{vice versa} is enough to
replace all the atmospheric carbon in 6 years.\cite{hard_science}
So, on time scales of interest, the partial pressure of
$CO_2$ in the atmosphere is proportional to the concentration
of $CO_2$ in the surface water of the oceans: they are in 
equilibrium. 

\section{The carbon cycle in the Amazon rain forest}
In the process of photosynthesis, a plant takes $CO_2$
from the air, breaks it up using solar energy, and 
releases $O_2$ back to the atmosphere. The carbon
attaches to water to form carbohydrates $(CH_2O)_n$, which
is what plants are made of. So, for every molecule of
oxygen released to the atmosphere, there remains one
atom of carbon in the plant.
What happens to the carbon
when the plant dies? There are three alternatives:
the carbon can (i) return to the atmosphere, (ii) accumulate on the ground,
or (iii) end up in the oceans.

The photosynthesis process can occur in both directions.
Examples of photosynthesis in reverse direction are 
combustion when we burn wood, respiration, and 
decomposition by aerobic bacteria. In these processes
the carbon of the plant becomes attached to oxygen
from the air, and is returned to the atmosphere as
$CO_2$. Note that no net oxygen is produced: the
oxygen released to the atmosphere during the plant growth is consumed
during the plant aerobic decay. 

The carbon can end up in the soil,
either as partially decomposed organic matter
(humus), as hydrocarbons or as coal.
If the plant dies in a medium that lacks oxygen, such
as in oceans, lakes, rivers, swamps or under volcanic ash,
the anaerobic bacteria have a chance to decompose the plant
(without the competition from aerobic bacteria).
The anaerobic bacteria produce methane (\textquotedblleft{swamp gas}"),
carbonic acids, bicarbonates, and humic acids. The methane
can escape to the atmosphere, or under enough pressure and
temperature, can polymerize into hydrocarbons or can be
crushed into coal.

The carbon can be washed down the rivers,
in the form of carbonic acids, bicarbonates, and humic acids which
are soluble in water. The rivers in the Amazon basin are dark due to the
humic acids.

The Amazon rain forest has humus only in the first few centimeters of
soil. So we neglect alternative (ii).
To quantify alternative (iii), we took samples of water of several
tributaries of the Amazon river (Napo Alto, Napo Bajo, San Francisco and
Tiputini) and measured the concentration of bicarbonate
(0.71 mol/m$^3$) and carbonic acid (0.11 mol/m$^3$).
Assuming that these concentrations are typical of the Amazon river,
and multiplying by the yearly discharge of the river, we obtain
the carbon discharged to the Atlantic ocean by the Amazon river:
50 million (metric) tonnes per year. In total, the Amazon basin fixes about
10 billion brute tonnes of carbon per year. So, about 99.5\% of this
carbon returns to the atmosphere due to the aerobic decomposition
of organic matter, while only about 0.5\% is washed down the
rivers into the Atlantic ocean. 

It is interesting to compare these numbers with the 
world energy-related release of carbon dioxide from the 
consumption of oil, gas and coal: 24 billion tonnes
in 2004. This corresponds to 6.5 billion 
tonnes of carbon per year. So the Amazon basin can 
transfer, from the atmosphere to the oceans, 
one year's worth of carbon dioxide
emission in 130 years!

\section{Conclusions}
With some degree of confidence, we arrive at the following
conclusions.
\begin{enumerate}
\item
Ever since we have accurate measurements of the temperature
of the atmosphere, \textit{i.e.} since 1702,
we observe a global increase of temperature of about 0.4$^0$C
per century. This warming does not accelerate in the second
half of the XX century. In fact, we see no statistically 
significant global warming since 1940.\cite{atmosphere, global-warming}
We observe no correlation of the global temperature with
the consumption of oil, coal and gas, or with population. 
The atmospheric concentrations of $CO_2$ begins to
increase around 1940\cite{atmosphere} when oil consumption takes off, 
yet we observe no corresponding
increase of the slope of global warming.\cite{atmosphere, global-warming,
CO2_observations} 
Some studies suggest that this is due to a balance between the
heating effect of $CO_2$ and the cooling effects of aerosols and
deforestation.\cite{hard_science, deforestation}
From the data it is concluded that the
accumulated effect of humankind on the global temperature 
until 1990 is
not statistically significant: $0.0 \pm 0.1 ^0$C.\cite{global-warming}
So, the observed global warming since 1702 is part of the natural variability
of the climate as we pull out of the \textquotedblleft{little ice age}"
three centuries ago.
\item
The pre-industrial concentration of $CO_2$ in the atmosphere was
$\approx 290$ ppm in 1800. It has since increased to 375 ppm in 2004.\cite{maunaloa}
There is convincing evidence that this increase is due to burning of
fossil fuels (and forrests, and the manufacture of cement).
\item
Detailed models estimate that the changes of atmospheric composition since
pre-industrial times up to 1995 produce a change of temperature of
$1.3 \pm 0.44$K due to all greenhouse gases (about half of this is
from $CO_2$), $-1.0 \pm 0.6$K due to aerosols\cite{hard_science},
and $\approx -0.4$K due to the change of albedo caused by
deforestation\cite{deforestation}.
So we do not know if human activity heats or cools the Earth.
As indicated above, no significant net change has been observed.
\item
The increase of the concentration of $CO_2$ in the atmosphere accurately tracks
the burning of fossil fuels (and forrests). Two thirds of the carbon burned
ends up in the atmosphere as $CO_2$, and the remaining third ends up dissolved in
the surface layer of the oceans (the time constant for this dissolution is about
6 years). The fraction of excess carbon in the atmosphere should decrease
to one third in about 50 years as more carbon is incorporated into the
oceans (as $CO_2$, $HCO_3^-$, $CO_3^{--}$, $CaCO_3$), and in the land
and marine biota.
The last third of the excess $CO_2$ in the atmosphere is expected
to return to the deep oceans and become buried as elemental carbon
and as carbonates with several time constants ranging from hundreds to thousands of
years.\cite{hard_science}
\item
Burning all of the remaining economically viable reserves of oil, gas and
coal over the next 150 years or so will approximately double the pre-industrial
atmospheric concentration of $CO_2$. The global warming due to a doubling
of the concentration of $CO_2$ is expected to increase the average surface air
temperature by 1.3K to 4.8K.\cite{hard_science} 
The increase of temperature is expected to be
higher than average at higher latitudes
(and lower than average at low latitudes). The heating effect of $CO_2$ is obvious
when we look at the Earth emission spectra shown in Figure \ref{earthradiation}.
The range of results is large, so global warming
is a difficult problem that does not seem to converge:
as the models become more complex, the uncertainties appear to
increase!
\item
Ice core samples indicate that $CO_2$ concentration and
temperature are well correlated: 
a change of temperature by 9.5$^0$C corresponds to
a change in $CO_2$ concentration 
from 195ppm to 270ppm.\cite{atmosphere} 
Does the change of $CO_2$ concentration cause the change of temperature
(by the greenhouse effect),
or does the change of temperature cause the change of $CO_2$ concentration
(due to the temperature dependence of the solubility of carbon in the oceans)?
Due to the greenhouse effect, an increase of $CO_2$ concentration
from 195ppm to 270ppm would cause an increase in temperature
of $\approx 0.8^0$C, much too small to account for the observed change
in temperature. Conversely, an increase of $9.5^0$C of the temperature of the
upper 1000m of the oceans would increase the atmospheric 
concentration of $CO_2$ by $\approx 44$ppm
(depending on the ocean pH). So it is plausible that the changes of temperature of the
oceans \textbf{caused} the pre-historic changes of atmospheric $CO_2$ concentration.
\item
Data suggests that large volcanic explosions can trigger transitions
from glacial to interglacial climates, by covering ice with 
volcanic ash, thereby changing the equilibrium temperature of the Earth 
as seen in Table \ref{emissivities}.
\item
Photosynthesis in the Amazon basin
fixes $\approx 10$ billion tonnes of carbon per year.
However, aerobic
decay releases almost all of this carbon back to the atmosphere as $CO_2$.
The balance, $\approx 50$ million tonnes of carbon per year,
is discharged to the Atlantic Ocean in the form of carbonic acids, 
bicarbonates, and humic acids. 

\end{enumerate}

\end{document}